\documentclass[twocolumn]{aastex63}

\usepackage{silence} 
\usepackage{times}
\usepackage{enumitem}
\usepackage{graphicx}
\usepackage{amsmath}
\usepackage{bm}
\usepackage{graphics}
\usepackage{url}
\usepackage{epsfig}
\usepackage{wrapfig}
\usepackage{ulem}
\usepackage{verbatim}
\usepackage{float}
\usepackage{lastpage}
\usepackage{longtable}

\usepackage{natbib}

\makeatletter
\define@key{Gin}{resolution}{\pdfimageresolution=#1\relax}
\makeatother

\providecommand{\apjl}{ApJ}

\providecommand{\apjs}{ApJS}
\providecommand{\aap}{A\&A}

\def\simlt{\mathrel{\hbox{\rlap{\hbox{\lower4pt\hbox{$\sim$}}}\hbox{$<$}}}}
\def\simgt{\mathrel{\hbox{\rlap{\hbox{\lower4pt\hbox{$\sim$}}}\hbox{$>$}}}}

\def\hst{{\it HST}}

\def\I{\,\textsc{i}}
\def\II{\,\textsc{ii}}

\def\V{\,\textsc{v}}

\def\arcsec{$^{\,\prime\prime}$}

\def\arcdeg{\degr}

\usepackage{longtable}
\usepackage{amssymb}

\shorttitle{SN\,2016gkg Progenitor and Follow up}
\shortauthors{Kilpatrick et~al.}

\begin{document}

\title[SN\,2016gkg Progenitor and Follow up]{Updated Photometry of the Yellow Supergiant Progenitor and Late-time Observations of the Type IIb Supernova 2016gkg}

\newcommand{\NU}{\affiliation{Center for Interdisciplinary Exploration and Research in Astrophysics (CIERA), Northwestern University, Evanston, IL 60208, USA}}
\newcommand{\UCSC}{\affiliation{Department of Astronomy and Astrophysics, University of California, Santa Cruz, CA 95064, USA}}
\newcommand{\carnegie}{\affiliation{The Observatories of the Carnegie Institution for Science, 813 Santa Barbara St., Pasadena, CA 91101, USA}}
\newcommand{\JHU}{\affiliation{Department of Physics and Astronomy, Johns Hopkins University, 3400 North Charles Street, Baltimore, MD 21218, USA}}
\newcommand{\STScI}{\affiliation{Space Telescope Science Institute, 3700 San Martin Drive, Baltimore, MD 21218, USA}}

\author[0000-0002-5740-7747]{Charles D. Kilpatrick}
\NU

\author[0000-0003-4263-2228]{David A. Coulter}
\UCSC

\author[0000-0002-2445-5275]{Ryan J. Foley}
\UCSC

\author[0000-0001-6806-0673]{Anthony L. Piro}
\carnegie

\author[0000-0002-4410-5387]{Armin Rest}
\JHU\STScI

\author[0000-0002-7559-315X]{C\'esar Rojas-Bravo}
\UCSC

\author{Matthew R. Siebert}
\UCSC

\begin{abstract}

We present {\it Hubble Space Telescope} (\hst) observations of the type IIb supernova (SN) 2016gkg at 652, 1698, and 1795~days from explosion with the Advanced Camera for Surveys (ACS) and Wide Field Camera 3 (WFC3).  Comparing to pre-explosion imaging from 2001 obtained with the Wide Field Planetary Camera 2, we demonstrate that SN\,2016gkg is now fainter than its candidate counterpart in the latest WFC3 imaging, implying that the counterpart has disappeared and confirming that it was the SN progenitor star.  We show the latest light curve and Keck spectroscopy of SN\,2016gkg, which implies that SN\,2016gkg is declining more slowly than the expected rate for ${}^{56}$Co decay during its nebular phase.  We find that this emission is too luminous to be powered by other radioisotopes, thus we infer that SN\,2016gkg is entering a new phase in its evolution where it is powered primarily by interaction with circumstellar matter.  Finally, we re-analyze the progenitor star spectral energy distribution and late-time limits in the context of binary evolution models and including emission from a potential companion star and find that all companion stars would be fainter than our limiting magnitudes.

\end{abstract}

\keywords{stars: evolution --- supernovae: general --- supernovae: individual (SN~2016gkg)}

%%%%%%%%%%%%%%%%%%%%%%%%%%%%%%%%%%∂%%%%%%%%%%%%%%%%%%%%%%%%%%%%%%%%%%%%
\section{INTRODUCTION}\label{sec:introduction}

Supernova (SN) 2016gkg is a type IIb SN \citep[SN\,IIb, with broad, transient lines of hydrogen;][]{filippenko97} that exploded in NGC~613 on 20 September 2016 \citep{kilpatrick17:16gkg}.  Similar to other nearby SNe\,IIb with deep pre-explosion imaging \citep[SNe~1993J, 2008ax, 2011dh, and 2013df;][]{aldering+94,crockett+08,maund+11,vandyk+14}, its counterpart was identified in archival {\it Hubble Space Telescope} (\hst) data as a yellow supergiant \citep{kilpatrick17:16gkg,tartaglia+16,Bersten18}.  This stellar classification is consistent with the expectation that SNe\,IIb have depleted hydrogen envelopes and thus hotter photospheric temperatures than the red supergiant (RSG) progenitor stars of type II-P SNe \citep{falk+73,falk+78,smartt09}.  However, the pathway through which these yellow supergiants shed their hydrogen envelopes remains ambiguous; the two most common explanations are enhanced radiative or eruptive mass loss \citep[e.g.,][]{Langer94,Maeder00,Heger03,Smith14} or interaction with and Roche-Lobe overflow onto a close binary companion \citep{Woosley+94,izzard2004,yoon+17}.  Additional constraints on the pre-explosion evolution, circumstellar environment, or possible companion star of SN\,2016gkg can therefore help to disentangle the pre-explosion evolution for SN~IIb progenitor stars generally.

SN\,2016gkg was also characterized by a rapid rise and luminous, ultraviolet-bright early peak in its light curve \citep{kilpatrick17:16gkg,Arcavi16,tartaglia+16,Bersten18}, which has been interpreted as shock cooling emission from the hydrogen envelope of its progenitor star.  This interpretation is consistent with observations of other well-observed SNe\,IIb with distinct early light curve peaks \citep[e.g., SN\,2011dh;][]{Arcavi12,bersten12}.  When compared with models that describe the rapid evolution as cooling in the outer envelope \citep{rabinak+11,nakar14,sapir16,piro17,Piro21}, photometric data can be used to independently constrain the size and mass of the hydrogen envelope.  For SN\,2016gkg, this radius was constrained as 30--150~$R_{\odot}$ \citep[see, e.g., fig. 4 in][]{Arcavi16}, depending primarily on the shock cooling model and envelope structure assumed.

This indirect constraint on the progenitor star mass can be directly compared to estimates from direct detection of the pre-explosion counterpart in {\it Hubble Space Telescope} ({\it HST}) Wide Field Planetary Camera 2 (WFPC2) data.  In \citet{kilpatrick17:16gkg}, comparison between the pre-explosion photometry and binary star models from BPASS \citep{eldridge+17} implied a progenitor star mass of around $15~M_{\odot}$.  Subsequent analyses of the same data yielded estimates of 15--20~$M_{\odot}$ \citep{tartaglia+16} and 19.5~$M_{\odot}$ \citep{Bersten18}, with both estimates invoking close binary models to explain the photometry and low hydrogen envelope mass in the progenitor star.  However, subsequent analyses using Modules for Experiments in Stellar Astrophysics (MESA) models suggest that for certain mass loss prescriptions there are both single-star and binary systems that can reproduce the pre-explosion photometry \citep{Sravan18,Gilkis21}.

Follow-up imaging of SN\,2016gkg can reveal whether the pre-explosion counterpart has disappeared, implying that it was indeed dominated by the progenitor star \citep[similar to the SN\,2011dh progenitor;][]{vandyk+13}, or if there is residual flux consistent with the pre-explosion source, implying that SN\,2016gkg exploded from a fainter star.  In addition, deep imaging, especially in bluer optical bands or the ultraviolet, may reveal the presence of a companion star at the site of SN\,2016gkg, similar to the putative companion of the type IIb SN 1993J \citep{maund+04,fox+14} and potentially 2011dh \citep{folatelli+14,maund+15}.  This is critical for analysis of the pre-explosion WFPC2 data, which have inferior resolution compared with later Advanced Camera for Surveys (ACS) and Wide Field Camera 3 (WFC3) imaging and may reveal blended sources near the SN\,2016gkg counterpart suggested in previous analyses \citep[e.g.,][]{tartaglia+16}.

Here we present follow-up {\it HST} and Swope imaging of SN\,2016gkg, spanning 32--1795~days from explosion, as well as late-time spectroscopy.  We present new photometry of the pre-explosion source using the improved relative astrometry of SN\,2016gkg provided by {\it HST} imaging at 652 and 1795~days post-explosion and updated analysis of that progenitor source.  We also analyze the light curve of SN\,2016gkg using the new late-time photometry as well as constraints on the progenitor system of SN\,2016gkg in the context of new photometry and limits on a companion star to the SN\,2016gkg progenitor.

As in \citet{kilpatrick17:16gkg}, we adopt a Tully-Fisher distance to NGC~613 of $D_{L}=26.4\pm5.3$~Mpc \citep{nasonova+11}.  We assume a line-of-sight Milky Way reddening of $E(B-V)=0.017$~mag \citep{Schlafly11}.  We further assume that there is additional line-of-sight extinction to SN\,2016gkg based on the relatively weak Na\I~D line observed in its spectra at the redshift of NGC~613 \citep{Arcavi16,tartaglia+16}, which combined with the conversions of \citet{Poznanski12} suggests emission from SN\,2016gkg was reddened by $E(B-V)=0.09\substack{+0.06\\-0.03}$~mag.

\section{Observations}\label{sec:observations}

\subsection{{\it HST} Imaging}

\begin{figure*}
    \centering
    \includegraphics[width=\textwidth]{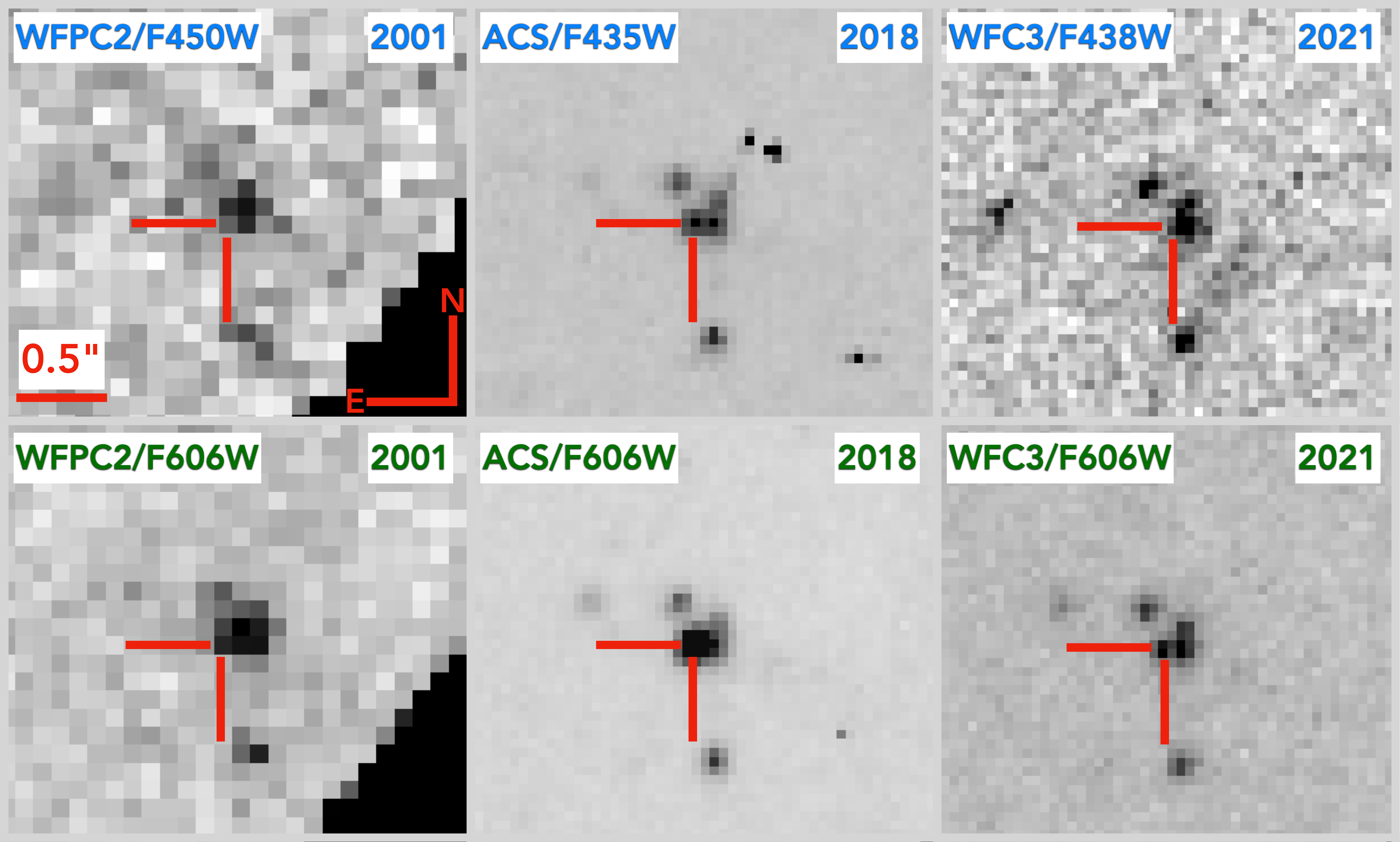}
    \caption{{\it HST} WFPC2, ACS, and WFC3 imaging of the site of SN\,2016gkg.  We highlight the location of SN\,2016gkg as determined from the ACS/WFC F606W image (bottom middle) from 2018, at which time the SN exhibited $m_{\rm F606W}=22.79\pm0.01$~mag (AB).  The counterpart appears fainter in WFC3 F438W and F606W imaging from 2021 (left top and bottom) than coincident pre-explosion emission from 2001 as observed in WFPC2 F450W and F606W (right top and bottom).}
    \label{fig:images}
\end{figure*}

We analyzed all {\it Hubble Space Telescope} ({\it HST}) covering the site of SN\,2016gkg, including the pre-explosion WFPC2 imaging from 2001, ACS imaging from 2018, and WFC3 imaging from 2021.  Following methods described in \citet{Kilpatrick21}, we downloaded all frames and aligned and drizzled them using the python-based code {\tt hst123} \citep{hst123}.  Using the drizzled ACS F606W frame as a reference, we performed photometry in each epoch using {\tt dolphot}.  We used standard {\tt dolphot} parameters for each epoch as described in {\tt hst123}.  All individual epochs and bands were drizzled separately, and we show all of the F450W, F435W, and F438W (roughly $B$-band) and F606W (roughly $V$-band) in \autoref{fig:images} and \autoref{tab:hst}.

SN\,2016gkg is clearly detectable in the post-explosion F606W frames at the site of the pre-explosion counterpart, and appears to fade significantly from 2018 to 2021 as expected for an evolved SN\,IIb.  By performing forced photometry at the site of this source in the remaining images, we are able to update and improve the photometric uncertainties on the counterpart detected in the WFPC2 data compared with \citet{kilpatrick17:16gkg}, which is also presented in \autoref{tab:hst}.

\begin{table}
    \centering
    \footnotesize
    \begin{tabular}{cccccc}
\hline
Epoch     & MJD         &      Instrument & Filter &   Magnitude & Uncertainty \\
(days)    &             &                 &        &    (mag)    & (mag)       \\ \hline\hline
$-$5482.29 & 52142.01 &  WFPC2      & F450W           &     24.37 &  0.22 \\
$-$5482.28 & 52142.02 &  WFPC2      & F606W           &     24.40 &  0.18 \\
$-$5482.27 & 52142.03 &  WFPC2      & F814W           &     24.43 &  0.20 \\
648.55     & 58302.89 &  ACS        & F435W           &     24.04 &  0.07 \\
648.62     & 58302.96 &  ACS        & F606W           &     22.59 &  0.01 \\
648.69     & 58303.03 &  ACS        & F814W           &     22.89 &  0.01 \\
1689.42    & 59348.86 &  WFC3       & F275W           &  $>$26.04 &  ---  \\
1785.90    & 59445.82 &  WFC3       & F606W           &     25.10 &  0.07 \\
1785.91    & 59445.83 &  WFC3       & F438W           &     26.61 &  0.27 \\ \hline
    \end{tabular}
    \caption{Pre- and Post-explosion {\it HST} photometry of the counterpart to SN\,2016gkg.  Epoch is given in rest-frame days relative to the explosion date of 2016 Sep 20.165 \citep{kilpatrick17:16gkg}.  All magnitudes are on the AB system \citep{Oke83}.}
    \label{tab:hst}
\end{table}

\subsection{Swope Imaging}\label{sec:swope}

\begin{figure}
    \centering
    \includegraphics[width=0.49\textwidth]{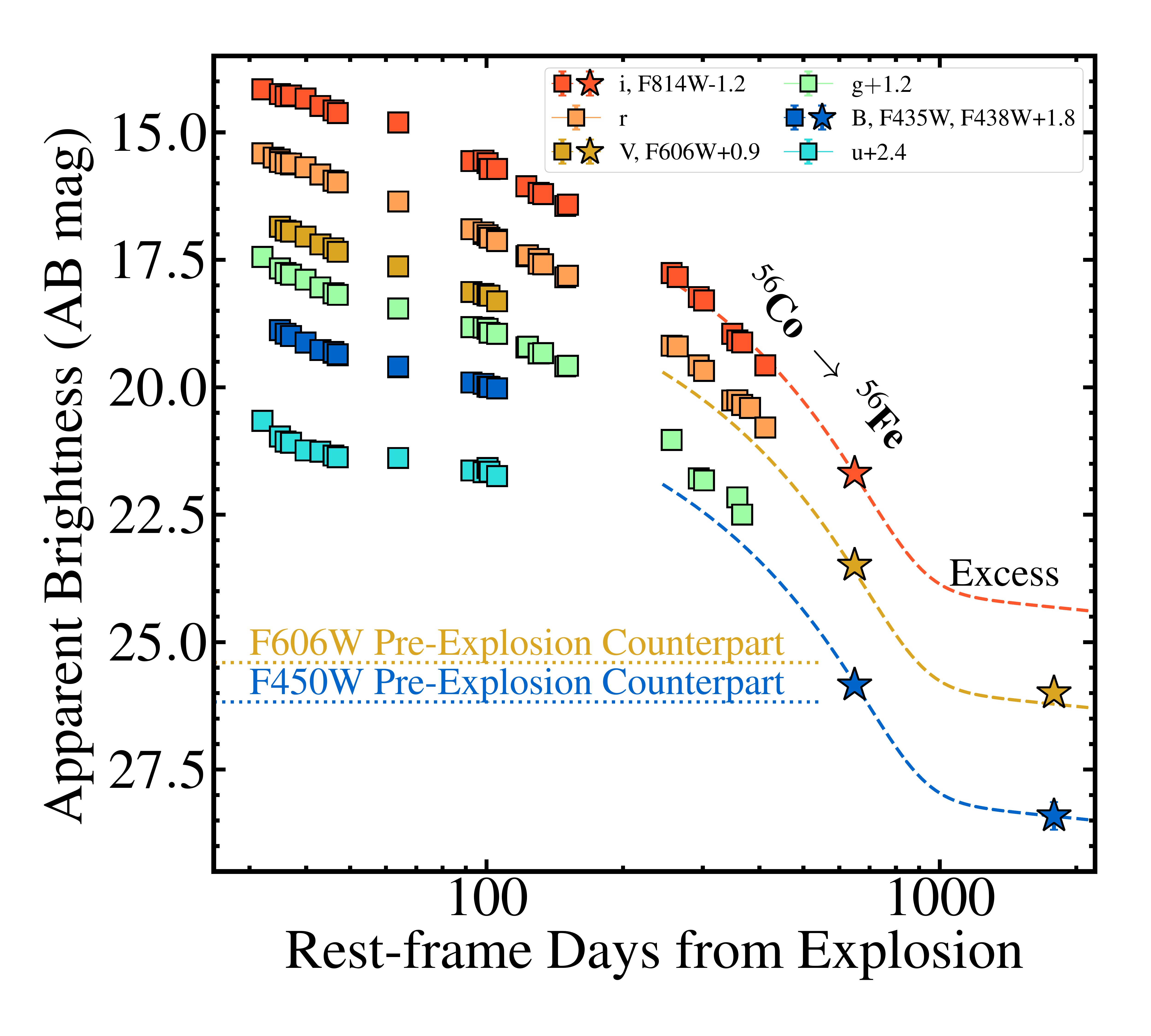}
    \caption{The combined $uBVgri$ Swope (squares) and F435W, F438W, F606W, and F814W {\it HST} (stars) light curve of SN\,2016gkg comprising photometry from +32 to +1795 days from explosion (32--1786 rest-frame days).  We compare F435W and F438W to $B$-band, F606W to $V$-band, and F814W to $i$-band.  For comparison, we show a simulated light curve powered by ${}^{56}$Co$\rightarrow{}^{56}$Fe decay (dashed line) to compare to the photometry at $>$250 rest-frame days.  However, in the 1795 day epoch, there appears to be excess emission (solid line) beyond the predicted emission from this model, which we model here using a circumstellar interaction model assuming a wind with $\dot{M}=$3--4$\times10^{-7}~M_{\odot}~\text{yr}$ and $v_{w}=$150--200~km~s$^{-1}$.}
    \label{fig:lightcurve}
\end{figure}

We observed SN\,2016gkg in $uBVgri$ bands from 22 October 2016 to 8 November 2017 with the Swope 1m telescope at Las Campanas Observatory, Chile.  Following standard procedures in the {\tt photpipe} imaging and photometry package \citep{rest+05} and as described in \citet{Kilpatrick18:16cfr}, all Swope data were corrected for bias and flat-fielded using calibration frames obtained in the same instrumental configuration as the SN\,2016gkg imaging.  We astrometrically aligned each Swope image to a common frame using 2MASS astrometric standards in the same field as SN\,2016gkg \citep{Skrutskie06}.  Finally, we regridded each image to correct for geometric distortion using {\tt SWarp} \citep{swarp} and performed photometry in the individual frames with {\tt DoPhot} \citep{schechter+93}.  The $BVgri$ data were photometrically calibrated using Pan-STARRS DR2 photometric standards in the same field as SN\,2016gkg and calibrated to the Swope natural system using SuperCal \citep{scolnic+15}.  The $u$-band data were calibrated using SkyMapper $u$-band standards \citep{skymapper}.  As SN\,2016gkg is relatively isolated and much brighter than its {\it HST} counterpart in all Swope data, we did not perform image subtraction on these images but instead use the unsubtracted frames to obtain final photometry, which is shown in \autoref{tab:swope} and \autoref{fig:lightcurve}.

\subsection{Keck/LRIS Spectroscopy}\label{sec:lris}

\begin{figure}
    \centering
    \includegraphics[width=0.49\textwidth]{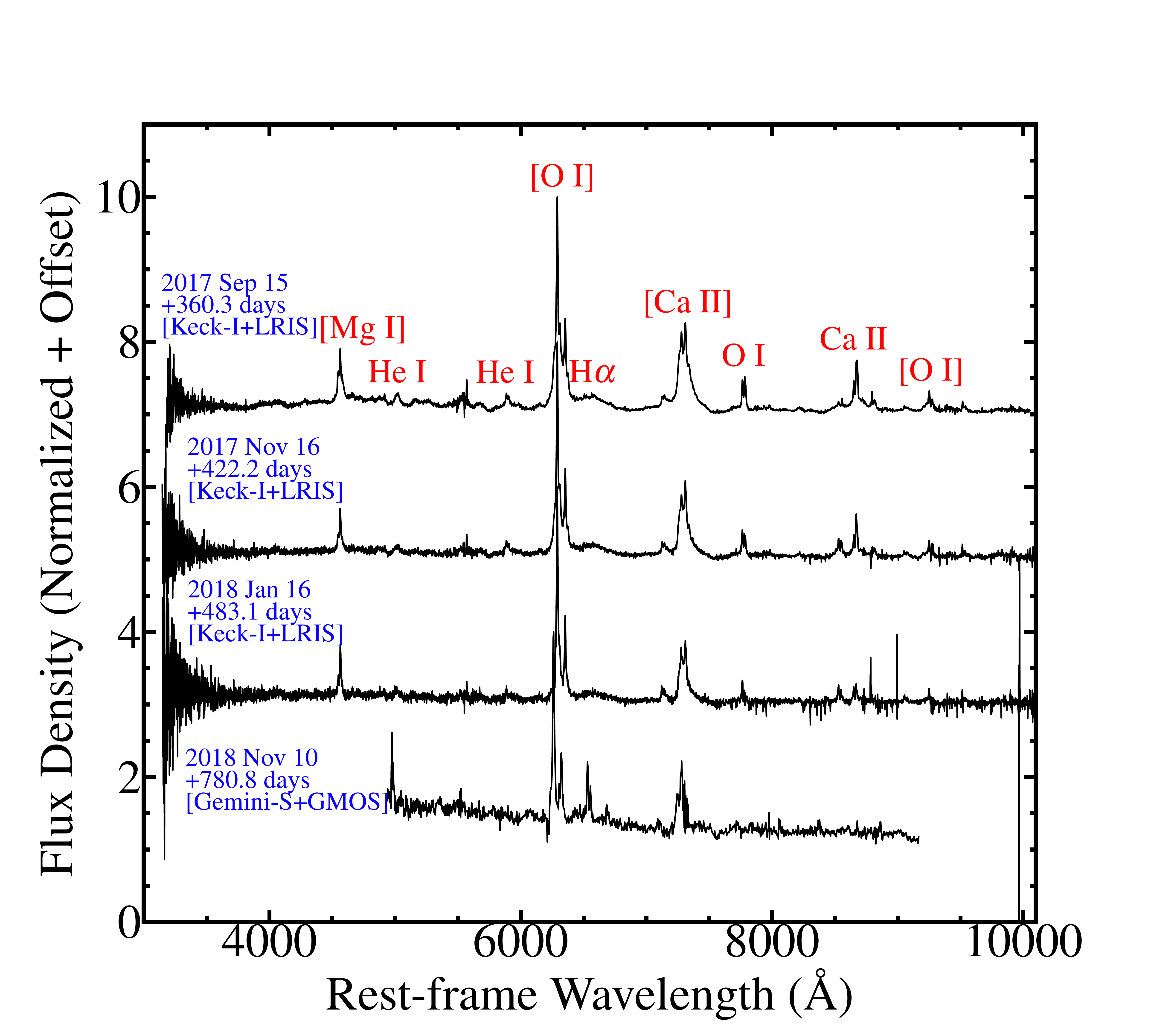}
    \caption{Our nebular spectra of SN\,2016gkg from Keck/LRIS as described in \autoref{sec:lris}.  We highlight lines of [Mg\I] $\lambda$4571, [O\I] $\lambda\lambda$6300, 6364, H$\alpha$, and [Ca\II] $\lambda\lambda$7292, 7324.  For comparison, we show the latest Gemini-S/GMOS spectrum from \citet{Kuncarayakti20}, which shows that H$\alpha$ emission is enhanced as the SN evolves.}
    \label{fig:spectra}
\end{figure}

We observed SN\,2016gkg with the Low Resolution Imaging Spectrometer on the Keck-I 10m telescope on 2017 September 15, 2018 November 16, and 2018 January 16 UT.  Each epoch was observed with the B600/4000 grism and R400/8500 grating in conjunction with the D560 dichroic and the 1.0\arcsec\ wide slit.  We reduced these data using standard {\tt IRAF}\footnote{IRAF is distributed by the National Optical Astronomy Observatory, which is operated by the Association of Universities for Research in Astronomy (AURA) under a cooperative agreement with the National Science Foundation.} and {\tt python} routines to correct for bias, flat-fielding, flux calibration, and telluric absorption.  All spectra are shown in \autoref{fig:spectra}.  As in \citet{Kuncarayakti20}, we detect emission lines of [Mg\I] $\lambda$4571, [O\I] $\lambda\lambda$6300, 6364, H$\alpha$, and [Ca\II] $\lambda\lambda$7292, 7324.  For comparison, we also plot the spectrum of SN\,2016gkg from Gemini-S+GMOS and obtained on 2018 November 10 presented in \citet{Kuncarayakti20}.  This final spectrum exhibits stronger emission in H$\alpha$, highlighting that SN\,2016gkg was likely entering a phase of enhanced interaction with circumstellar matter (CSM).

\section{Analysis}\label{sec:analysis}

\subsection{\hst\ Photometry of SN\,2016gkg and Its Progenitor Star}\label{sec:sed}

We note that SN\,2016gkg declines significantly in F435W/F438W and F606W from 2018 to 2021.  Indeed, SN\,2016gkg is fainter than its pre-explosion counterpart in F438W (comparing to F450W) and F606W.  We infer from this decline that the SN\,2016gkg pre-explosion counterpart has disappeared and is in fact the progenitor star.

\begin{figure}
    \centering
    \includegraphics[width=0.49\textwidth]{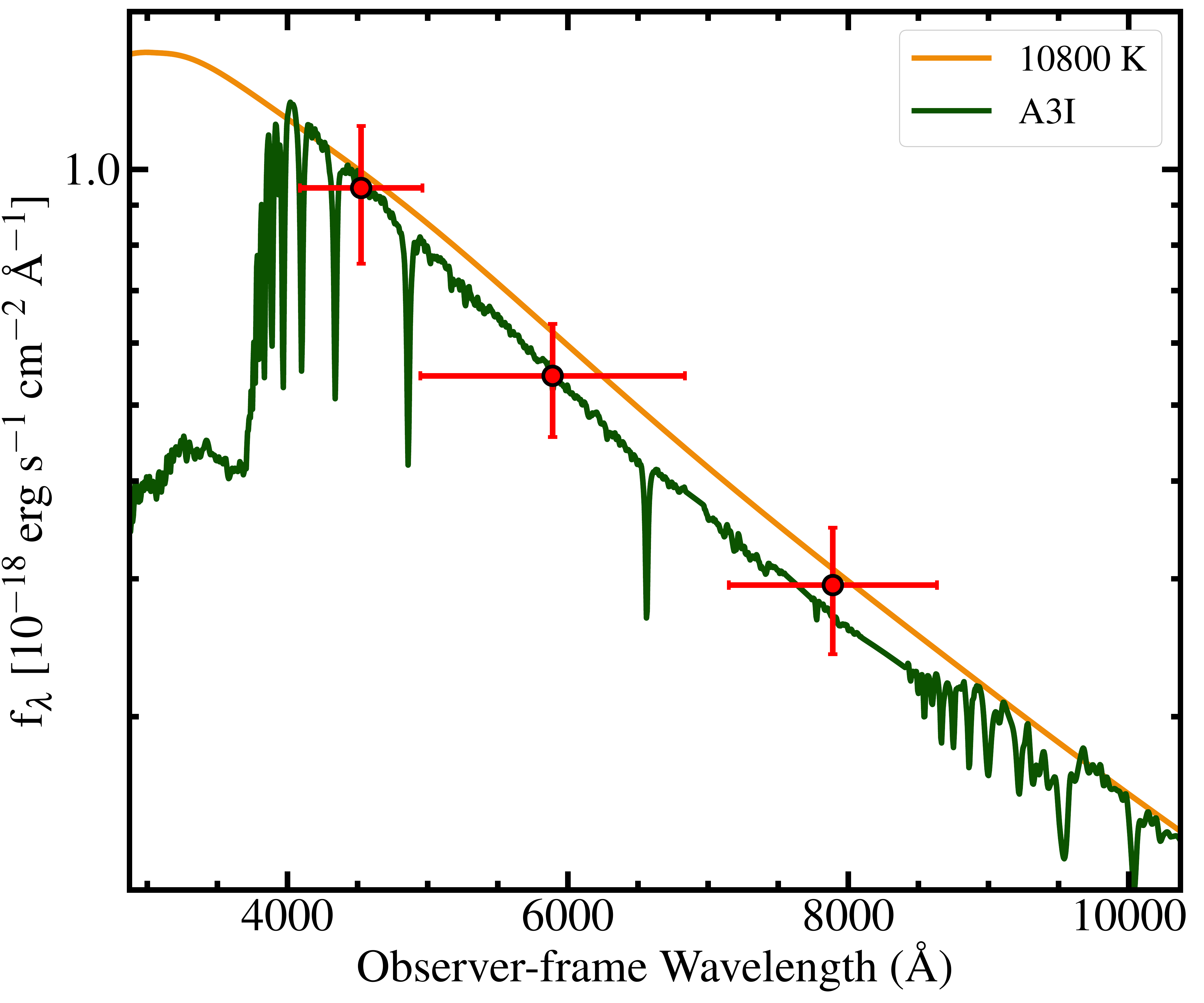}
    \caption{Spectral energy distribution (SED) of the SN\,2016gkg pre-explosion counterpart from F450W, F606W, and F814W (red points) with vertical error bars including the measurement uncertainty and horizontal error bars indicating the effective filter width.  We overplot two models used to describe the overall SED; a blackbody with $T_{\rm eff}=10,800$~K and a stellar SED for a A3I star derived from \citet{pickles+10} models as described in \autoref{sec:sed}.}
    \label{fig:sed}
\end{figure}

Using the updated progenitor star photometry from \autoref{tab:hst}, we use a blackbody model and stellar spectral energy distributions (SEDs) from \citet{pickles+98} to infer the effective temperature and luminosity of that star.  Following methods in \citet{Kilpatrick21}, we fit the pre-explosion photometry using a Monte Carlo Markov Chain approach and find that the progenitor star SED is best fit by a blackbody with $T_{\rm eff}=10800\pm1000$~K and $\log(L/L_{\odot})=4.7\pm0.5$ as shown in \autoref{fig:sed}.  The uncertainty in the latter quantity is dominated by the uncertainty in the distance modulus.  Such a blackbody photosphere implies a radius of 60$\pm$30~$R_{\odot}$.

From our fits using \citet{pickles+98} SEDs, we find the best-fitting model is a A3 star with $T_{\rm eff}=8900\substack{+1100\\-700}$~K and $\log(L/L_{\odot})=4.5\pm0.5$.  Given the lower temperature, this source would have a marginally more extended photospheric radius of $70\pm30~R_{\odot}$.  The difference in effective temperature from the blackbody models arises because we are probing the Rayleigh-Jeans tail of the SED in both cases and the stellar SED models provide a more realistic, absorbed quasi-photosphere and UV suppression at $>$4000\AA\ due to metal lines (\autoref{fig:sed}) that also matches the progenitor photometry at cooler temperatures.  Therefore, we consider the A3 star temperature and implied luminosity to be a better match to the progenitor detections.  Taking the endpoint luminosity as an indicator for the initial mass of the progenitor star, this would place SN\,2016gkg at 10$\substack{+2\\-1}$~$M_{\odot}$, in agreement with the expectations for source B in \citet{tartaglia+16}.

\subsection{Late-time Photometric Evolution of SN\,2016gkg and Its Spectral Energy Distribution}

SN\,2016gkg appears to decline following the radioactive decay rate of ${}^{56}$Co as shown in \autoref{fig:lightcurve}.  As in \citet{Prentice18}, we model this phase of the light curve assuming that SN\,2016gkg initially produced 0.085~$M_{\odot}$ of ${}^{56}$Ni and assuming its light curve is powered primarily by radioactive decay from 250--652~days post-explosion.  This assumption is supported by the overall light curve shape at this phase as well as the lack of strong spectroscopic features due to interaction with CSM at $<$500~days \citep[\autoref{fig:spectra} and][]{Kuncarayakti20}.

From 652 to 1795~days post-explosion, there is significant flattening in the light curve compared with predictions from ${}^{56}$Co decay.  We infer that some other source of emission dominated SN\,2016gkg in the most recent epoch.  At the same time, the color $m_{\rm F435W}-m_{\rm F606W}=1.45\pm0.07$~mag at 652~days is consistent with the color $m_{\rm F438W}-m_{\rm F606W}=1.51\pm0.28$~mag at 1795 days.  This finding is consistent with expectations for a nebular SN, where the emission line dominated spectrum undergoes a ``freeze out'' at late times \citep[$>$1~year as in][]{Maeda07,Jerkstrand11,Jerkstrand14} as cooling due to emission lines is balanced by the heating rate due to the shock and radioactive species.

\begin{figure}
    \centering
    \includegraphics[width=0.49\textwidth]{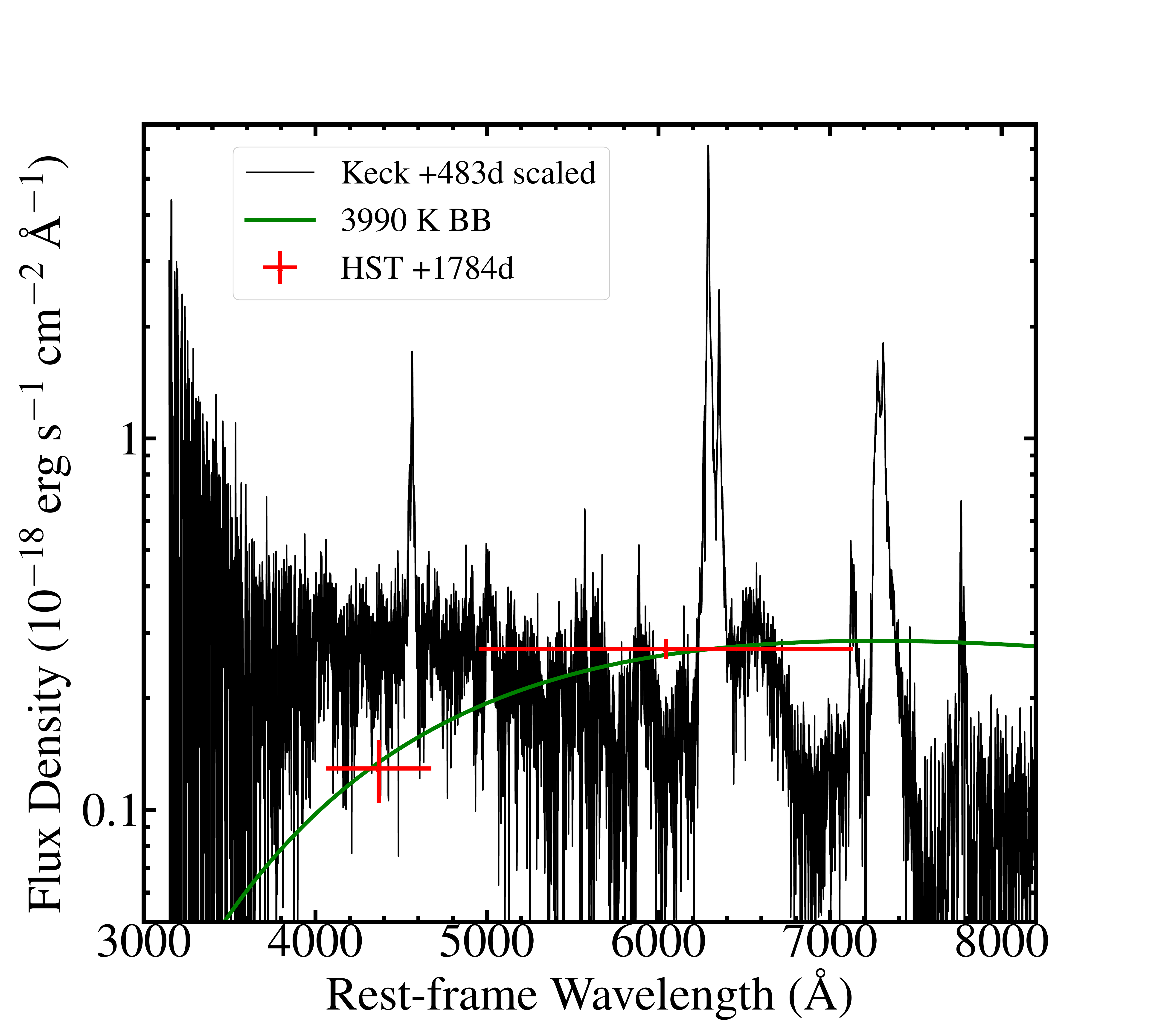}
    \caption{The \hst\ photometry from SN\,2016gkg at +1786 rest-frame days (red) as shown in \autoref{fig:lightcurve} compared with a scaled Keck/LRIS spectrum of SN\,2016gkg from +483~days (black) and the best-fitting blackbody spectrum with $T_{\rm eff}=3990$~K (green).  The scaled Keck spectrum corresponds to a much bluer color than the late-time \hst\ photometry, implying that there is significant spectral evolution out to this late phase, such as H$\alpha$ emission as in \autoref{fig:spectra}, or there is some other source of emission dominating SN\,2016gkg.}
    \label{fig:late}
\end{figure}

In order to test whether the late-time photometry is consistent with coming from SN\,2016gkg, we show the latest F438W and F606W \hst\ photometry in \autoref{fig:late} with comparisons to a blackbody spectral energy distribution similar to those used for the pre-explosion photometry in \autoref{sec:sed} and our +483~day Keck/LRIS spectrum, which covers $>$90\% of the enclosed energy for both \hst\ bands.  If the underlying spectral source is a blackbody, such as from a coincident star, the implied temperature and luminosity are $T_{\rm eff}=3990$~K and $\log(L/L_{\odot})=4.3$.  On the other hand, the photometry is not well fit to the Keck/LRIS spectrum, with the spectrum suggesting $m_{\rm F438W}-m_{\rm F606W}=0.54$~mag (AB) compared with 1.51~mag from the photometry.  This implies that there is either some other source or significant spectral evolution at +1795~days compared with $\sim1300$~days earlier.  Based solely on the spectral evolution observed in \autoref{sec:lris}, we infer that there may be significantly more CSM interaction in the latest photometric observations.  Thus the H$\alpha$ feature, which is observable in F606W, dominates the emission profile.  However, we cannot rule out the possibility that some blended source or another emission mechanism contributing to the photometry.

To account for the excess emission above that of ${}^{56}$Co-decay, we first consider other radioactive species such as ${}^{57}$Co, ${}^{55}$Fe, and ${}^{44}$Ti.  Integrating the emission in F438W and F606W at 1795~days and accounting for extinction and the distance $26.4$~Mpc to SN\,2016gkg, we find the total pseudo-bolometric $B+V$-band emission is $2.6\pm0.5\times10^{38}$~erg~s$^{-1}$ (including the uncertainty for distance).  Under the assumption that all of this emission is due to various radioisotopes with 100\% thermalization efficiency, we can place a lower limit on the total mass of that radioisotopes assuming it accounts for all of the SN luminosity.  Again assuming 100\% thermalization and a rest-frame time $t$, a radioisotope with atomic mass $m_{A}$, decay constant $\lambda_{A}$, mean energy per decay $q_{A}$, and initially ($t=0$) synthesized with a total mass $M_{A}$ will produce a luminosity

\begin{equation}
    L(t) = \frac{\lambda_{A} M_{A} q_{A}}{m_{A}} e^{-\lambda_{A} t}\label{eqn:bateman}.
\end{equation}

\noindent We consider ${}^{57}$Co ($\lambda=2.56\times10^{-3}$~day$^{-1}$, $m=56.9$~u $q=836.0$~keV), ${}^{55}$Fe ($\lambda=6.93\times10^{-4}$~day$^{-1}$, $m=54.9$~u, $q=231.4$~keV), and ${}^{44}$Ti ($\lambda=3.16\times10^{-5}$~day$^{-1}$, $m=44.0$~u, $q=3920.8$~keV) to compare to \autoref{eqn:bateman} \citep{Firestone99}.  For ${}^{44}$Ti, we have assumed that the daughter isotope ${}^{44}$Sc, with a half-life of 3.9~hr, decays effectively instantaneously and added its energy per decay with that of its parent.  Assuming SN\,2016gkg emitted $2.6\pm0.5\times10^{38}$~erg~s$^{-1}$ at a rest-frame time of 1785.9~days, we find minimum initial masses of 0.03~$M_{\odot}$ (${}^{57}$Co), 0.01~$M_{\odot}$ (${}^{55}$Fe), and 0.004~$M_{\odot}$ (${}^{44}$Ti).  For the former two radioisotopes, these would imply extremely large mass ratios to the primary iron-peak isotope ${}^{56}$Ni that are not favored by nucleosynthesis in core-collapse SNe \citep[e.g., in calculations by][]{Curtis19,Andrews20}.  Even for ${}^{44}$Ti, which has been observed in the Galactic supernova remnant Cassiopeia A with an implied initial mass $1.5\times10^{-4}~M_{\odot}$ \citep{Grefenstette16} and in SN\,1987A with an initial mass $1.5\times10^{-4}~M_{\odot}$ \citep{Boggs15}, SN\,2016gkg would require at least $25\times$ more of this radioisotope than has been observed for other core-collapse SNe.  We therefore find it more likely that there is some other energy source powering the late-time emission.

We next consider that the counterpart may be a light echo arising from the SN light scattered by a local dust sheet into our line of sight.  Light echoes have been observed for a few extragalactic SNe \citep[e.g., SN~2006X and 2014J;][]{Wang10,Crotts15} in addition to Galactic SN remnants such as Cassiopeia A and Tycho \citep{Rest08}.  For unresolved light echoes from sufficiently thick dust sheets, the light echo SED resembles the luminosity-weighted average of their complete time-varying SED, which will be dominated by the SN SED around peak light.  As an approximation, we estimate that the $m_{\rm F435W}-m_{\rm F606W}$ color of a SN\,2016gkg light echo would be its color around the second $V$-band maximum, which was about $B-V=0.4$~mag \citep[AB;][]{tartaglia+16}.  Including emission from the initial but shorter first peak, this color would be even bluer.  Thus the redder $m_{\rm F435W}-m_{\rm F606W}=1.51$~mag counterpart we see is unlikely to be from a light echo.

Motivated by observations of possible CSM interaction in other SN\,IIb light curves \citep[e.g., SNe\,1993J, 2011dh, and 2013df in][]{Maeda15,smith+17,Kundu19,Maund19}, we model the excess emission using a CSM interaction model for a $\rho\propto r^{-2}$ wind profile due to a constant mass loss rate $\dot{M}$ and wind speed $v_{w}$.  We use a relatively simple prescription for this emission with the bulk luminosity following the analytical model in \citet[][their equation 1, implying luminosity $\propto t^{-3}$]{smith+10c} and no color evolution from the previous epochs.  Assuming a wind speed of $\approx$150--200~km~s$^{-1}$, corresponding to the escape speed of a star with a terminal mass 7--10~$M_{\odot}$ and radius 90~$R_{\odot}$, the excess corresponds to a mass-loss rate of $\dot{M}=$3--4$\times10^{-7}~M_{\odot}~\text{yr}$.  This is a relatively low mass-loss rate for a $10~M_{\odot}$ star \citep[e.g., compared with][]{smith+04,vink11,Smith14}, but it is a plausible scenario considering the systematic uncertainties in our light curve model.  Detailed modeling of the complete radio, X-ray, and optical light curve of SN\,2016gkg can provide a more complete picture of that system's pre-explosion mass-loss history and what progenitor scenarios it matches. However, the excess need not arise entirely from CSM interaction but may be partly from a static source coincident with SN\,2016gkg.  We investigate the excess optical emission in the context of possible companion models below.

\subsection{The Spectral Energy Distribution of SN\,2016gkg at +1786~days and Limits on a Companion Star to the SN\,2016gkg Progenitor}

In \citet{kilpatrick17:16gkg}, the authors used publicly available binary stellar evolution models from the Binary Population and Spectral Synthesis code \citep[BPASS;][]{eldridge+17} to consider scenarios in which one star in a binary system could produce SN\,2016gkg.  We repeat that analysis here but including the additional constraints on the presence of a companion star either assuming that the excess emission observed from SN\,2016gkg represents a companion or an upper limit on a companion star if it is still dominated by the SN itself.  We also include upper limits in the ultraviolet filter F275W observed at 1689~days in \autoref{tab:hst}.

Among the BPASS models in v2.2, the best-fitting binary stellar endpoint for our data involves a 10+3~$M_{\odot}$ binary where the primary terminates with a final mass $M_{f}=2.2~M_{\odot}$ and a envelope hydrogen mass of $0.07~M_{\odot}$.  At the time the primary star explodes, the secondary star is still a B-type main sequence star with $m>30$~mag at the distance of NGC~613 in optical and ultraviolet bands.  This would imply that the late-time counterpart does not correspond to the companion star and our limits are unconstraining for any binary scenarios.

Even in the most extreme cases allowed by our photometry of the SN\,2016gkg progenitor star, the secondary star can have at most an initial mass of $7~M_{\odot}$ with $m_{V}\approx30.7$~mag at the time the primary explodes.  This reinforces our finding above that the spectral energy distribution of the late-time emission is more consistent with SN\,2016gkg itself.

\section{Discussion \& Conclusions}\label{sec:discussion}

The improved astrometric alignment between SN\,2016gkg and its pre-explosion counterpart over ground-based adaptive optics imaging \citep{kilpatrick17:16gkg,tartaglia+16} and short-exposure \hst\ imaging in which the SN is nearly saturated \citep{Bersten18} enables a novel analysis of its progenitor system.  It is clear from the superior resolution of the late-time \hst\ imaging that the WFPC2 counterpart is indeed split into multiple components as hypothesized by \citet{tartaglia+16}, and the counterpart corresponds to their source B.  Our revised photometry is in good agreement with their original analysis, although we have improved the photometric precision by using the position of SN\,2016gkg provided by the late-time ACS and WFC3 imaging.  Our updated estimates of the luminosity of this source further suggest that SN\,2016gkg exploded from a relatively compact ($\approx$70~$R_{\odot}$) yellow supergiant star with an initial mass of 10$\substack{+2\\-1}$~$M_{\odot}$.  These results agree with most of the early-time shock cooling modeling \citep{Arcavi16,kilpatrick17:16gkg,piro17,tartaglia+16,Piro21}, especially those using the \citet{piro15} and \citet{sapir16} that tend to lead to smaller radii.

Investigating the late-time evolution of SN\,2016gkg, the light curve initially follows the ${}^{56}$Co decay rate, which is also suggested by the lack of H$\alpha$ emission in the nebular spectra of \citet{Kuncarayakti20}.  The relative brightnesses of SN\,2016gkg in F435W, F606W, and F814W (\autoref{tab:hst}) are consistent with expectations that the SED of SN\,2016gkg is dominated by [Mg\I] $\lambda$4571, [O\I] $\lambda\lambda$6300, 6364, and [Ca\II] $\lambda\lambda$7292, 7324 at around 652~days.  It is therefore unlikely that there is any significant contribution from CSM interaction at these phases, although detailed radio and X-ray observations could reveal an underlying shock interaction with a surrounding wind or other medium.

However, at 1786 rest-frame days, there appears to be a flattening in the light curve implying that it is no longer ${}^{56}$Co-powered.  The emission is too bright to be any other radioisotope typically found in core-collapse SNe such as ${}^{44}$Ti.  At this very late phase, it is possible that CSM interaction in the shock begins to dominate the optical radiation provided by radioactive decay, especially as the latest optical spectrum of SN\,2016gkg showed an enhancement in H$\alpha$ emission, likely due to an interaction with H-rich material. While it is possible that there is some static source contributing to the SN\,2016gkg counterpart at this phase, the spectral energy distribution of that emission is consistent with earlier spectra of SN\,2016gkg, and so we find it more likely that the SN still dominates the \hst\ photometry.  Moreover, our initial search of BPASS models that are consistent with the counterpart do not reveal any possible companion stars that can simultaneously reproduce the 1786~day counterpart and the SN\,2016gkg progenitor source. Additional follow up of SN\,2016gkg is needed reveal if the source continues to decline in brightness consistent with CSM interaction or if there remains a persistent source of emission.

\startlongtable
\begin{deluxetable}{cccc}
\tablewidth{0pt}
\tablecaption{Swope Photometry of SN\,2016gkg\label{tab:swope}}
\tablewidth{0pt}
\tablehead{
\colhead{MJD} &
\colhead{Filter} &
\colhead{Magnitude} &
\colhead{Uncertainty} \\
&
&
(mag) &
(mag)
}
\startdata
57686.31 & $B$ & 17.08 & 0.01 \\
57687.33 & $B$ & 17.15 & 0.01 \\
57688.34 & $B$ & 17.19 & 0.01 \\
57691.20 & $B$ & 17.32 & 0.01 \\
57694.33 & $B$ & 17.47 & 0.01 \\
57697.29 & $B$ & 17.52 & 0.04 \\
57697.30 & $B$ & 17.51 & 0.01 \\
57698.30 & $B$ & 17.57 & 0.03 \\
57698.31 & $B$ & 17.55 & 0.01 \\
57715.25 & $B$ & 17.81 & 0.05 \\
57715.25 & $B$ & 17.79 & 0.01 \\
57744.18 & $B$ & 18.11 & 0.02 \\
57750.08 & $B$ & 18.13 & 0.02 \\
57752.14 & $B$ & 18.19 & 0.01 \\
57753.14 & $B$ & 18.20 & 0.05 \\
57753.14 & $B$ & 18.19 & 0.01 \\
57757.14 & $B$ & 18.22 & 0.01 \\
57683.31 & $g$ & 16.24 & 0.01 \\
57686.30 & $g$ & 16.47 & 0.01 \\
57687.32 & $g$ & 16.56 & 0.01 \\
57688.33 & $g$ & 16.59 & 0.01 \\
57691.19 & $g$ & 16.69 & 0.01 \\
57694.32 & $g$ & 16.84 & 0.01 \\
57697.31 & $g$ & 16.96 & 0.01 \\
57698.32 & $g$ & 16.99 & 0.01 \\
57715.27 & $g$ & 17.25 & 0.01 \\
57744.19 & $g$ & 17.62 & 0.01 \\
57750.07 & $g$ & 17.63 & 0.01 \\
57752.13 & $g$ & 17.65 & 0.01 \\
57753.16 & $g$ & 17.73 & 0.01 \\
57757.12 & $g$ & 17.75 & 0.01 \\
57774.13 & $g$ & 18.02 & 0.02 \\
57775.12 & $g$ & 18.00 & 0.02 \\
57782.09 & $g$ & 18.14 & 0.01 \\
57785.06 & $g$ & 18.13 & 0.01 \\
57801.03 & $g$ & 18.39 & 0.02 \\
57803.04 & $g$ & 18.38 & 0.02 \\
57908.41 & $g$ & 19.83 & 0.02 \\
57946.37 & $g$ & 20.59 & 0.08 \\
57954.43 & $g$ & 20.63 & 0.03 \\
58010.32 & $g$ & 20.96 & 0.05 \\
58019.38 & $g$ & 21.30 & 0.05 \\
57683.31 & $i$ & 15.35 & 0.01 \\
57686.30 & $i$ & 15.45 & 0.01 \\
57687.32 & $i$ & 15.49 & 0.01 \\
57688.33 & $i$ & 15.48 & 0.01 \\
57691.19 & $i$ & 15.54 & 0.01 \\
57694.32 & $i$ & 15.68 & 0.01 \\
57697.31 & $i$ & 15.77 & 0.01 \\
57698.32 & $i$ & 15.82 & 0.01 \\
57715.27 & $i$ & 16.01 & 0.01 \\
57744.19 & $i$ & 16.76 & 0.01 \\
57750.07 & $i$ & 16.75 & 0.01 \\
57752.13 & $i$ & 16.80 & 0.01 \\
57753.16 & $i$ & 16.92 & 0.01 \\
57757.12 & $i$ & 16.92 & 0.01 \\
57774.13 & $i$ & 17.26 & 0.02 \\
57782.09 & $i$ & 17.39 & 0.01 \\
57785.06 & $i$ & 17.41 & 0.01 \\
57801.03 & $i$ & 17.64 & 0.02 \\
57803.04 & $i$ & 17.62 & 0.02 \\
57908.41 & $i$ & 18.96 & 0.02 \\
57916.43 & $i$ & 19.04 & 0.03 \\
57946.36 & $i$ & 19.43 & 0.03 \\
57954.42 & $i$ & 19.50 & 0.03 \\
58001.25 & $i$ & 20.15 & 0.05 \\
58010.31 & $i$ & 20.28 & 0.04 \\
58019.37 & $i$ & 20.32 & 0.04 \\
58065.23 & $i$ & 20.77 & 0.09 \\
57683.30 & $r$ & 15.41 & 0.02 \\
57683.31 & $r$ & 15.41 & 0.01 \\
57685.20 & $r$ & 15.49 & 0.02 \\
57686.29 & $r$ & 15.54 & 0.02 \\
57686.29 & $r$ & 15.53 & 0.01 \\
57686.29 & $r$ & 15.58 & 0.01 \\
57687.32 & $r$ & 15.61 & 0.02 \\
57688.32 & $r$ & 15.63 & 0.02 \\
57688.32 & $r$ & 15.61 & 0.01 \\
57691.19 & $r$ & 15.68 & 0.01 \\
57694.32 & $r$ & 15.83 & 0.01 \\
57697.32 & $r$ & 15.95 & 0.01 \\
57698.32 & $r$ & 15.98 & 0.01 \\
57715.27 & $r$ & 16.36 & 0.01 \\
57744.19 & $r$ & 16.90 & 0.01 \\
57750.06 & $r$ & 16.99 & 0.03 \\
57750.07 & $r$ & 16.96 & 0.01 \\
57752.12 & $r$ & 17.02 & 0.01 \\
57753.16 & $r$ & 17.07 & 0.01 \\
57757.12 & $r$ & 17.10 & 0.04 \\
57757.12 & $r$ & 17.13 & 0.01 \\
57774.13 & $r$ & 17.43 & 0.01 \\
57775.11 & $r$ & 17.41 & 0.05 \\
57775.12 & $r$ & 17.41 & 0.05 \\
57782.08 & $r$ & 17.48 & 0.05 \\
57782.09 & $r$ & 17.58 & 0.01 \\
57785.06 & $r$ & 17.59 & 0.01 \\
57801.02 & $r$ & 17.84 & 0.02 \\
57803.04 & $r$ & 17.81 & 0.01 \\
57908.40 & $r$ & 19.18 & 0.02 \\
57916.43 & $r$ & 19.20 & 0.02 \\
57946.36 & $r$ & 19.57 & 0.03 \\
57954.41 & $r$ & 19.68 & 0.03 \\
58001.24 & $r$ & 20.26 & 0.06 \\
58010.30 & $r$ & 20.25 & 0.04 \\
58019.36 & $r$ & 20.35 & 0.03 \\
58034.24 & $r$ & 20.41 & 0.09 \\
58065.22 & $r$ & 20.79 & 0.08 \\
57683.32 & $u$ & 18.26 & 0.03 \\
57686.30 & $u$ & 18.57 & 0.03 \\
57687.32 & $u$ & 18.66 & 0.03 \\
57688.33 & $u$ & 18.69 & 0.03 \\
57691.19 & $u$ & 18.84 & 0.03 \\
57694.33 & $u$ & 18.86 & 0.04 \\
57697.30 & $u$ & 18.94 & 0.03 \\
57698.31 & $u$ & 18.97 & 0.03 \\
57715.26 & $u$ & 18.99 & 0.04 \\
57744.18 & $u$ & 19.23 & 0.06 \\
57750.07 & $u$ & 19.27 & 0.03 \\
57752.13 & $u$ & 19.18 & 0.03 \\
57753.15 & $u$ & 19.26 & 0.04 \\
57757.13 & $u$ & 19.34 & 0.04 \\
57686.31 & $V$ & 15.95 & 0.01 \\
57687.33 & $V$ & 16.02 & 0.01 \\
57688.34 & $V$ & 16.05 & 0.01 \\
57691.20 & $V$ & 16.14 & 0.01 \\
57694.33 & $V$ & 16.30 & 0.01 \\
57697.30 & $V$ & 16.37 & 0.01 \\
57698.31 & $V$ & 16.44 & 0.01 \\
57715.26 & $V$ & 16.73 & 0.01 \\
57744.18 & $V$ & 17.23 & 0.01 \\
57750.08 & $V$ & 17.27 & 0.01 \\
57752.14 & $V$ & 17.30 & 0.01 \\
57753.15 & $V$ & 17.30 & 0.01 \\
57757.14 & $V$ & 17.41 & 0.01
\enddata
\tablecomments{All photometry is on the AB magnitude system.}
\end{deluxetable}

\bigskip\bigskip\bigskip
\noindent {\bf ACKNOWLEDGMENTS}
\smallskip
\footnotesize

We would like to thank I. Arcavi and A. Gilkis for helpful comments related to this manuscript.
This work includes data obtained with the Swope Telescope
at Las Campanas Observatory, Chile, as part of the Swope Time Domain Key Project (PI Piro; Co-Is Burns, Cowperthwaite, Dimitriadis, Drout, Foley, French, Holoien, Hsiao, Kilpatrick, Madore, Phillips, and Rojas-Bravo).
The {\it Hubble Space Telescope} imaging presented in this manuscript come from programs GO-9042 (PI Smartt), GO-15272 (PI Folatelli), GO-16287 (PI Lyman), and SNAP-16179 (PI Filippenko).
Based on observations made with the NASA/ESA {\it Hubble Space Telescope}, obtained from the data archive at the Space Telescope Science Institute. STScI is operated by the Association of Universities for Research in Astronomy, Inc. under NASA contract NAS 5-26555.
Some of the data presented herein were obtained at the W. M. Keck Observatory, which is operated as a scientific partnership among the California Institute of Technology, the University of California, and NASA. The Observatory was made possible by the generous financial support of the W. M. Keck Foundation.
The authors wish to recognize and acknowledge the very significant cultural role and reverence that the summit of Maunakea has always had within the indigenous Hawaiian community.  We are most fortunate to have the opportunity to conduct observations from this mountain.

\facilities{{\it HST} (WFPC2, ACS, WFC3), Keck (LRIS), Swope (Direct)}

\software{{\tt dolphot} \citep{dolphot},
          {\tt DoPhot} \citep{schechter+93},
          {\tt hst123} \citep{hst123},
          {\tt photpipe} \citep{rest+05},
          {\tt SWarp} \citep{swarp}}

% REFERENCES
\bibliography{2016gkg}

\end{document}